\documentclass[fleqn]{llncs}
\usepackage{xspace}
\usepackage{url}
\usepackage{xcolor}
\usepackage{soul}
\usepackage{amsfonts}
\usepackage{amsmath}
\usepackage{amssymb}
\usepackage{z-eves}
\usepackage{booktabs}
\usepackage{tabularx}
\usepackage{colortbl}
\usepackage{tikz}
\usetikzlibrary{positioning,patterns,shapes,arrows,shadows,backgrounds,fit}
\usepackage[figuresleft,counterclockwise]{rotating}
\usepackage{rotating}
\usepackage{multicol}
\usepackage{comment}

\newcommand{\setlog}{$\{log\}$\xspace}
\newcommand{\SET}{\mathcal{SET}}
\newcommand{\BR}{\mathcal{BR}}
\newcommand{\RIS}{\mathcal{RIS}}

\newcommand{\LCARD}{\mathcal{CARD}}   
\newcommand{\LINT}{\mathcal{INTV}}    
\newcommand{\LIA}{\mathcal{LIA}}   

\newcommand{\F}{\mathcal{F}} 
\newcommand{\C}{\mathcal{C}} 
\newcommand{\Pred}{\mathcal{P}} 
\newcommand{\Cl}{\mathcal{H}} 
\newcommand{\Q}{\mathcal{Q}} 

\newcommand{\es}{\{\}}
\newcommand{\false}{\mathtt{false}}
\newcommand{\true}{\mathtt{true}}

\renewcommand{\And}{\mathbin{{\text{\footnotesize\&}}}}
\renewcommand{\Cup}{\mathtt{un}}
\renewcommand{\Cap}{\mathtt{inters}}
\newcommand{\Int}{\mathtt{int}}
\newcommand{\Cp}{\mathtt{cp}}
\newcommand{\In}{\mathbin{\mathtt{in}}}
\newcommand{\Disj}{\mathtt{disj}}
\newcommand{\Size}{\mathtt{size}}
\newcommand{\Dom}{\mathtt{dom}}
\newcommand{\Ran}{\mathtt{ran}}
\newcommand{\Inv}{\mathtt{inv}}
\newcommand{\Id}{\mathtt{id}}
\newcommand{\Oplus}{\mathtt{oplus}}
\newcommand{\Subseteq}{\mathtt{subset}}
\newcommand{\Comp}{\mathtt{comp}}

\newcommand{\Apply}{\mathtt{apply}}
\newcommand{\Pfun}{\mathtt{pfun}}

\newcommand{\Or}{\mathbin{\mathtt{or}}}
\newcommand{\Ncup}{\mathtt{nun}}
\newcommand{\Nin}{\mathbin{\mathtt{nin}}}
\newcommand{\Ris}{\mathtt{ris}}
\newcommand{\Neq}{\mathbin{\mathtt{neq}}}
\newcommand{\Forall}{\mathtt{foreach}}
\newcommand{\Implies}{\mathbin{\mathtt{implies}}}
\newcommand{\Neg}{\mathtt{neg}}
\newcommand{\Str}{\mathtt{str}}
\newcommand{\Etype}{\mathtt{etype}}
\newcommand{\Stype}{\mathtt{stype}}
\newcommand{\Dec}{\mathtt{dec}}
\newcommand{\ApplyTo}{\mathtt{applyTo}}
\newcommand{\Foplus}{\mathtt{foplus}}
\newcommand{\Exists}{\mathtt{exists}}

\def\comp{\mathrel{\raise 0.66ex\hbox{\oalign{\hfil%
        $\scriptscriptstyle\mathsf{o}$\hfil%
        \cr\hfil$\scriptscriptstyle\mathsf{9}$\hfil}}}}

\newcommand{\z}{$\epsilon$\xspace}

\newcommand{\hlg}[1]{\sethlcolor{green}\hl{#1}}
\newcommand{\hlc}[1]{\sethlcolor{cyan}\hl{#1}}

\title{An Automatically Verified Prototype of a Landing Gear System}

\author{Maximiliano Cristi\'a\inst{1} \and Gianfranco Rossi\inst{2}}

\institute{%
Universidad Nacional de Rosario and CIFASIS,
Rosario, Argentina \\
\email{cristia@cifasis-conicet.gov.ar}
\and
Universit\`a di Parma, Parma, Italy \\ \email{gianfranco.rossi@unipr.it}
}

\begin{document}

\maketitle

\begin{abstract}
In this paper we show how \setlog (read `setlog'), a Constraint Logic
Programming (CLP) language based on set theory, can be used as an automated
verifier for B specifications. In particular we encode in \setlog an Event-B
specification, developed by Mammar and Laleau, of the case study known as the
Landing Gear System (LGS). Next we use \setlog to discharge all the proof
obligations proposed in the Event-B specification by the Rodin platform. In
this way, the \setlog program can be regarded as an automatically verified
prototype of the LGS. We believe this case study provides empirical evidence on
how CLP and set theory can be used in tandem as a vehicle for program
verification.
\end{abstract}

\section{Introduction}\label{sec:introduction}

In the fourth edition of the ABZ Conference held in Toulouse (France) in 2014,
ONERA's Boniol and Wiels proposed a real-life, industrial-strength case study,
known as the Landing Gear System (LGS) \cite{DBLP:conf/asm/BoniolW14}. The
initial objectives of the proposal were ``to learn from experiences and
know-how of the users of the ASM, Alloy, B, TLA, VDM and Z methods'' and to
disseminate the use of verification techniques based on those methods, in the
aeronautic and space industries. The LGS was the core of an ABZ track of the
same name. The track received a number of submissions of which eleven were
published \cite{DBLP:conf/asm/BoniolW14}. Six of the published papers
approached the problem with methods and tools rooted in the B notation
\cite{Abrial00} (Event-B, ProB, Hybrid Event-B and Rodin). In this paper we
consider the article by Mammar and Laleau \cite{DBLP:conf/asm/MammarL14} and a
journal version \cite{DBLP:journals/sttt/MammarL17} as the starting point for
our work. In those articles, the authors use Event-B as the specification
language and Rodin \cite{DBLP:journals/sttt/AbrialBHHMV10}, ProB
\cite{Leuschel00} and AnimB\footnote{\url{http://www.animb.org}} as
verification tools.

The B method was introduced by Abrial \cite{Abrial00} after his work on the Z
notation \cite{Spivey00}. B is a formal notation based on state machines, set
theory and first-order logic aimed at software specification and verification.
Verification is approached by discharging proof obligations generated during
specification refinement. That is, the engineer starts with a first, abstract
specification and refines it into a second, less abstract specification. In
order to ensure that the refinement step is correct a number of proof
obligations must be discharged. This process is continued until an executable
program is obtained. Given that all refinements have been proved correct, the
executable program is correct by construction.

Discharging a proof obligation entails to perform a \emph{formal proof} (most
often) in the form of a \emph{mechanized proof}. A formal proof is a proof of a
mathematical theorem; a mechanized proof is a formal proof made by either an
interactive theorem prover or an automated theorem prover\footnote{In this
context the term `automated theorem prover' includes tools such as
satisfiability solvers.}. That is, mechanized proofs are controlled, guided or
verified by a program.  Unless there are errors in the verification software, a
mechanized proof is considered to be error-free because those programs are
supposed to implement only sound proof steps. The mechanization of proofs is
important for at least two reasons: errors cannot be tolerated in
safety-critical systems and mechanization enables the possibility of proof
automation which in turn reduces verification costs.

Our work starts by considering the Event-B specification
of the LGS developed by Mammar and Laleau. Event-B
\cite{Abrial:2010:MES:1855020} is a further development over B aimed at
modeling and reasoning about discret-event systems. The basis of B are
nonetheless present in Event-B: state machines, set theory, first-order logic,
refinement and formal proof. The Event-B specification developed by Mammar and
Laleau has an important property for us. They used the Rodin platform to write
and verify the model. This implies that proof obligations were generated by
Rodin according to a precise and complete algorithm
\cite{DBLP:journals/sttt/AbrialBHHMV10}.

In this paper we consider the LGS casy study and Mammar and Leleau's Event-B
specification as a benchmark for \setlog (read `setlog').  \setlog is a
Constraint Logic Programming (CLP) language and satisfiability solver based on
set theory. As such, it can be used as a model animator and as an automated
theorem prover.  Since \setlog is based on set theory, it should also be
a good candidate to encode Event-B (or classic B) specifications; since it
implements several decision procedures for set theory, it should be a good
candidate to automatically discharge proof obligations generated from
refinement steps of Event-B (or classic B) specifications.

Therefore, we proceed as follows: \emph{a)} Mammar and Leleau's Event-B
specification is encoded as a \setlog program; \emph{b)} all the proof
obligations generated by the Rodin platform are encoded as \setlog queries; and
\emph{c)} \setlog is used to automatically discharge all these queries. We say
`encode' and not `implement' due to the similarities between the \setlog
language and the mathematical basis of the Event-B language; however, the
encoding provides an implementation in the form of a prototype.

The contributions of this paper are the following:
\begin{itemize}
\item We provide empirical evidence on how CLP and set theory can be used
in tandem as a vehicle for program verification. More specifically, \setlog is
shown to work well in practice.
\item Given that the \setlog prototype of the LGS has been (mechanically
and automatically) proved to verify a number of properties, it can be regarded
as correct w.r.t. those properties.
\end{itemize}

The paper is structured as follows. In Sect. \ref{setlog} and \ref{usingsetlog}
we introduce \setlog by means of several examples. In particular, in Sect.
\ref{usingsetlog} we shown the \emph{formula-program duality} enjoined by
\setlog. Section \ref{encoding} presents the encoding of the Event-B
specification of the LGS in \setlog. The encoding of the proof obligations
generated by the Rodin tool is introduced in Sect. \ref{proofs}. Finally, we
discuss our approach in Sect. \ref{discussion} and give our conclusions in
Sect. \ref{concl}.

\section{\label{setlog}Overview of \setlog}

\setlog is a publicly available satisfiability solver and a declarative
set-based, constraint-based programming language implemented in Prolog
\cite{setlog}. \setlog is deeply rooted in the work on Computable Set Theory
\cite{10.5555/92143}, combined with the ideas put forward by the set-based
programming language SETL \cite{DBLP:books/daglib/0067831}.

\setlog implements various decision procedures for different theories on the
domain of finite sets and integer numbers. Specifically, \setlog implements:
a decision procedure for the theory of \emph{hereditarily finite sets}
($\SET$), i.e., finitely nested sets that are finite at each level of nesting
\cite{Dovier00}; a decision procedure for a very expressive fragment of the
theory of finite set relation algebras ($\BR$)
\cite{DBLP:journals/jar/CristiaR20,DBLP:conf/RelMiCS/CristiaR18}; a decision
procedure for the theory $\SET$ extended with
restricted intensional sets ($\RIS$) \cite{DBLP:journals/jar/CristiaR21a}; a
decision procedure for the theory
$\SET$ extended with cardinality constraints ($\LCARD$)  \cite{cristia_rossi_2021}; a
decision procedure for the latter extended with integer intervals ($\LINT$)
\cite{DBLP:journals/corr/abs-2105-03005}; and integrates an existing decision
procedure for the theory of linear integer arithmetic ($\LIA$). All these
procedures are integrated into a single solver, implemented in Prolog, which
constitutes the core part of the \setlog tool. Several in-depth empirical
evaluations provide evidence that \setlog is able to solve non-trivial problems
\cite{DBLP:journals/jar/CristiaR20,DBLP:conf/RelMiCS/CristiaR18,DBLP:journals/jar/CristiaR21a,CristiaRossiSEFM13};
in particular as an automated verifier of security properties
\cite{DBLP:journals/jar/CristiaR21,DBLP:journals/jar/CristiaR21b}.

\subsection{The theories}

\begin{figure}
\begin{center}
\begin{tikzpicture}
  [every node/.style={transform shape},font=\sffamily,
   rectangle,text centered,minimum width=2cm,minimum height=1cm]
\draw
 (0,0) node[draw,fill=gray!20] (ilp) {$\LIA$}
 (4,0) node [draw,minimum width=6cm,fill=gray!20] (set) {$\SET$}
 (1,1) node [draw,minimum width=4cm,fill=gray!20] (card) {$\LCARD$}
 (4.5,1) node [minimum width=1cm,fill=gray!20] {}
 (4,1) node [draw] (ris) {$\RIS(\mathcal{X})$}
 (5.5,1) node [minimum width=1cm,fill=gray!20] {}
 (6,1) node [draw] (ra) {$\BR$}
 (1,3) node [minimum width=4cm,fill=gray!20] {}
 (1,3) node [minimum width=4cm,pattern=dots] {}
 (5,3) node [minimum width=2cm,minimum height=3cm,fill=gray!20] {}
 (1,4) node [minimum width=4cm,fill=gray!20] {}
 (1,2) node [draw,minimum width=4cm,fill=gray!20] (int) {$\LINT$}
 (5,2.5) node [minimum width=4cm,
               minimum height=2cm, align=right, pattern=dots] {}
 (3,3) node [draw,minimum width=8cm,minimum height=3cm] {}
 (3,3) node [fill=white,
             minimum width=0cm,
            minimum height=0cm, inner sep=3pt] (arr) {ARRAY}
 (3,4) node [draw,minimum width=8cm,pattern=dots] {}
 (3,4) node [fill=white,
             minimum width=0cm,
             minimum height=0cm, inner sep=3pt] (list) {LIST}
 (9,2) node [fill=gray!20,
             minimum height=10pt] (dec) {{\tiny\ decidable}}
 (9,3) node [minimum height=10pt,
             draw] (undec) {{\tiny undecidable}}
 (9,4) node [pattern=dots,minimum height=10pt] {}
 (9,4) node [fill=white,minimum width=0cm,inner sep=0pt,
             minimum height=0cm] (wip) {{\tiny work in progress}};
\end{tikzpicture}
\end{center} \caption{\label{fig:stack}The stack of theories dealt with by
\setlog}
\end{figure}
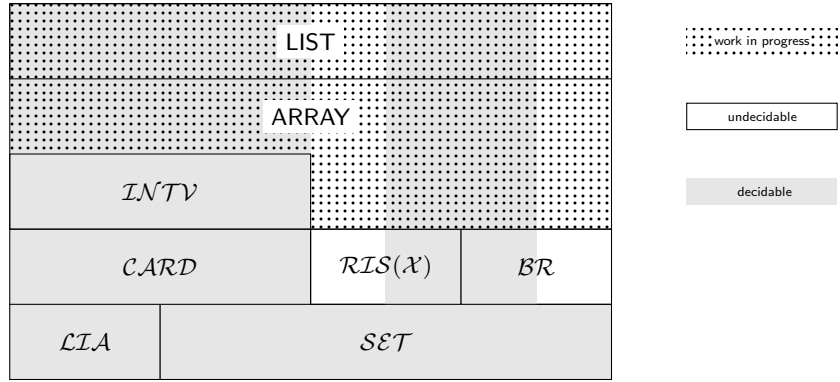

Figure \ref{fig:stack} schematically describes the stack of the first-order
theories supported by \setlog. The fact that theory $T$ is over  theory $S$
means that $T$ extends $S$. For example, $\LCARD$ extends both $\LIA$ and
$\SET$.

$\LIA$ provides integer linear arithmetic constraints
(e.g.,
$2*X + 5 \leq 3*Y - 2$) by means of Prolog's CLP(Q) library
\cite{DBLP:conf/cp/HolzbaurMB96}.
$\SET$ \cite{Dovier00} provides the Boolean algebra of hereditarily finite
sets; that is, it provides equality ($=$), union ($\Cup$), disjointness
($\Disj$), and membership ($\In$). In turn, these operators enable the
definition of other operators, such as intersection and relative complement, as
$\SET$ formulas. In all set theories, set operators are encoded as atomic
predicates, and are dealt with as constraints. For example, $\Cup(A,B,C)$ is a
constraint interpreted as $C = A \cup B$.
$\LCARD$ \cite{cristia_rossi_2021} extends $\SET$ and $\LIA$ by providing the
cardinality operator ($\Size$) which allows to link sets with integer
constraints\footnote{`$\And$' stands for conjunction ($\land$); see Sect.
\ref{formulas}.} (e.g.,
$\Size(A,W) \And W - 1 \leq 2*U + 1$).
$\RIS$ \cite{DBLP:journals/jar/CristiaR21a} extends $\SET$ by introducing the
notion of restricted intensional set (RIS) into the Boolean algebra. RIS are
finite sets defined by a property. For example, $x \in \{y:A | \phi(y)\}$,
where $A$ is a set and $\phi$ is a formula of a parameter theory $\mathcal{X}$.
$\BR$ \cite{DBLP:journals/jar/CristiaR20,DBLP:conf/RelMiCS/CristiaR18} extends
$\SET$ by introducing ordered pairs, binary relations and Cartesian products
(as sets of ordered pairs), and the operators of relation algebra lacking in
$\SET$---identity ($\Id$), converse ($\Inv$) and composition ($\Comp$). For
example\footnote{$[x,y]$ stands for the ordered pair $(x,y)$.},
$\Comp(R,S,\{[X,Z]\}) \And \Inv(R,T) \And [X,Y] \In T$. In turn, $\BR$ allows
the definition of relational operators such as domain, range and domain
restriction, as $\BR$ formulas.
$\LINT$ \cite{DBLP:journals/corr/abs-2105-03005} extends $\LCARD$ by
introducing finite integer intervals thus enabling the definition of several
non-trivial set operators such as the minimum of a set and the partition of a
set w.r.t. a given number.
\textsf{ARRAY} and \textsf{LIST} are still work in progress. They should
provide theories to (automatically) reason about arrays and lists from a set
theoretic perspective. For example, $A$ is an array of length $n$ iff is a
function with domain in the integer interval\footnote{$\Int(a,b)$ stands for
the integer interval $[a,b]$; see Sect. \ref{terms}.} $\Int(1,n)$. Then, in
\textsf{ARRAY} it is possible to define the predicate $\mathtt{array}(A,n)
\defs \Pfun(A) \And \Dom(A,\Int(1,n))$, where $\Pfun$ and $\Dom$ are predicates
definable in $\BR$. However, it is still necessary to study the decidability of
such an extension to $\BR+\LINT$ as, at the bare minimum, it requires for
$\Dom$ to deal with integer intervals.
The following subsections provide some clarifications on the syntax and
semantics of the logic
languages on which these theories are based on.

\subsection{\label{terms}Set terms}

The integrated constraint language offered by \setlog is a first-order
predicate language with terms of three sorts: terms designating sets (i.e.,
\emph{set terms}), terms designating integer numbers, and terms designating
ur-elements (including ordered pairs, written as $[x,y]$). Terms of either sort
are allowed to enter in the formation of set terms (in this sense, the
designated sets are hybrid), no nesting restrictions being enforced (in
particular, membership chains of any finite length can be modeled).

Set terms in \setlog can be of the following forms:
\begin{itemize}
\item A variable is a set term; variable names start with an uppercase letter.
\item $\{\}$ is the term interpreted as the empty set.
\item $\{t / A\}$, where $A$ is a set term and
$t$ is any term accepted by \setlog (basically,
any Prolog uninterpreted term, integers, ordered pairs, other set terms, etc.),
is called \emph{extensional set} and is interpreted as
$\{t\} \cup A$. As a notational convention, set
terms of the form $\{t_1 / \{t_2 \,/\, \cdots \{ t_n / t\}\cdots\}\}$ are
abbreviated as $\{t_1,t_2,\dots,t_n / t\}$, while $\{t_1 / \{t_2 \,/\, \cdots
\{ t_n / \es\}\cdots\}\}$ is abbreviated as $\{t_1,t_2,\dots,t_n\}$.
\item $\Ris(X \In A, \phi)$, where $\phi$ is any \setlog formula, $A$ is
a any set term among the first three, and $X$ is a bound variable local to the
$\Ris$ term, is called \emph{restricted intensional set} (RIS) and is
interpreted as $\{x : x \in A \land \phi\}$. Actually, RIS have a more complex
and expressive structure \cite{DBLP:journals/jar/CristiaR21a}.
\item $\Cp(A,B)$, where $A$ and $B$ are any set term among the first three, is interpreted
as $A \times B$, i.e., the Cartesian product between $A$ and $B$.
\item $\Int(m,n)$, where $m$ and $n$ are either integer constants or variables,
is interpreted as $\{x \in \num | m \leq x \leq n\}$.
\end{itemize}

Set terms can be combined in several ways: binary relations are hereditarily
finite sets whose elements are ordered pairs and so set operators can take
binary relations as arguments; RIS and integer intervals can be passed as
arguments to $\SET$ operators and freely combined with extensional sets.

\subsection{Set and relational operators}

\setlog implements a wide range of set and relational operators covering most
of those used in B. Some of the basic operators are provided as
\emph{primitive} constraints. For instance, $\Pfun(F)$ constrains $F$ to be a
(partial) function; $\Dom(F,D)$ corresponds to $\dom F = D$; $\Subseteq(A,B)$
corresponds to $A \subseteq B$; $\Comp(R,S,T)$ is interpreted as $T = R \comp
S$ (i.e., relational composition); and $\Apply(F,X,Y)$ is equivalent to
$\Pfun(F) \And [X,Y] \In F$.

A number of other set, relational and integer operators (in the form of
predicates) are defined as \setlog formulas, thus making it simpler for the
user to write complex formulas. Dovier et al. \cite{Dovier00} proved that the
collection of predicate symbols implementing
$\{=,\neq,\in,\notin,\cup,\parallel\}$ is sufficient to define constraints
for the set operators $\cap$, $\subseteq$ and
$\setminus$.
This result has been extended to binary relations
\cite{DBLP:journals/jar/CristiaR20} by showing that adding to the previous
collection the predicate symbols implementing $\{\id,\comp,{}^\smile\}$
is sufficient to define constraints for most of the classical relational
operators, such as $\dom$, $\ran$, $\dres$, $\rres$, etc. Similarly,
$\{=,\neq,\leq\}$ is sufficient to define $<$, $>$ and $\geq$. We call
predicates defined in this way \emph{derived constraints}.

\setlog provides also so-called \emph{negated
constraints}. For example, $\Ncup(A,B,C)$ is interpreted as $C \neq A \cup B$
and $\Nin$ corresponds to $\notin$---in general, a constraint beginning with
`$\mathtt{n}$' identifies a negated constraint. Most of these constraints are
defined as derived constraints in terms of the existing primitive constraints;
thus their introduction does not really require extending the constraint
language.

\subsection{\label{formulas}\setlog formulas}

Formulas in \setlog are built in the usual way by using the propositional
connectives (e.g., $\And$, $\Or$) between atomic constraints. More precisely,
\setlog formulas are defined according to the following grammar:
\begin{equation*}
 \F ::= \true | \false | \C | \F \And \F | \F \Or \F | \Pred |
 \Neg(\F) | \F \Implies \F | \Q(\F)
\end{equation*}
where: $\C$ is any \setlog atomic constraint; $\Pred$ is any atomic predicate
which is defined by a Horn clause of the form:
\begin{equation*}
 \Cl ::= \Pred | \Pred \text{ :- } \F
\end{equation*}
and the user-defined predicates possibly occurring in the body of the clause do
not contain any direct or indirect recursive call to the predicate in $\Pred$;
$\Q(\F)$ is the quantified version of $\F$, using either restricted universal
quantifiers (RUQ) or restricted existentially quantifiers (REQ). The definition
and usage of RUQ and REQ in \setlog will be discussed in Section
\ref{ssect:quantifiers}. It is worth noting that fresh variables occurring in
the body of a clause but not in its head are all implicitly existentially
quantified.


$\Neg$ computes the \emph{propositional} negation of its argument. In
particular, if $\F$ is an atomic constraint, $\Neg(\F)$ returns the
corresponding negated constraint. For example, $\Neg(x \In A \And z \Nin C)$
becomes $x \Nin A \Or z \In C$.
However, the result of $\Neg$ is not always correct because, in general, the
negated formula may involve existentially quantified variables, whose negation
calls into play (general) universal quantification that \setlog cannot handle
properly. Existentially quantified variables may appear in the bodies of
clauses or in the formula part of RIS and RUQ. Hence, there are cases where
\setlog users must manually compute the negation of some formulas.

The same may happen for some logical connectives, such as $\Implies$, whose
implementation uses the predicate $\Neg$.\footnote{Indeed, $F \Implies G$ is
implemented in \setlog as $\Neg(F) \Or G$.}

\subsection{Types in \setlog}
\setlog is an untyped formalism. This means that, for example, a set such as
$\{(x,y),1,`red`\}$ is accepted by the language. However, recently, a type
system and a type checker have been defined and implemented in \setlog. Typed
as well as untyped formalisms have advantages and disadvantages
\cite{DBLP:journals/toplas/LamportP99}. For this reason, \setlog users can
activate and deactivate the typechecker according to their needs.

\setlog types are defined according to the following grammar:
\begin{equation*}
\tau ::= \Int | \Str | Atom | \Etype([Atom,\dots,Atom]) | [\tau,\dots,\tau] | \Stype(\tau)
\end{equation*}
where $Atom$ is any Prolog atom other than $\Int$ and $\Str$. $\Int$
corresponds to the type of integer numbers; $\Str$ corresponds to the type of
Prolog strings; if $atom \in Atom$, then it defines the type given by the set
$\{atom\mathtt{?}t | t \text{ is a Prolog atom}\}$, where `$\mathtt{?}$'
is a functor of arity 2. $\Etype([c_1,\dots,c_n])$, with $2 \leq n$, defines
an enumerated type; $[T_1,\dots,T_n]$ with $2 \leq n$ defines the Cartesian
product of types $T_1,\dots,T_n$; and $\Stype(T)$ defines the powerset type of
type $T$. As can be seen, this type system is similar to B's.

When in typechecking mode, all variables and user-defined predicates must be
declared to be of a precise type. Variables are declared by means of the
$\Dec(V,\tau)$ constraint, meaning that variable $V$ is of type $\tau$. In this
mode, every \setlog atomic constraint has a polymorphic type much as in the B
notation. For example, in $\Comp(R,S,T)$ the type of the arguments must be
$\Stype([\tau_1,\tau_2])$, $\Stype([\tau_2,\tau_3])$ and
$\Stype([\tau_1,\tau_3])$, for some types $\tau_1, \tau_2, \tau_3$,
respectively.

Concerning user-defined predicates, if the head of a predicate is
$\texttt{p(}X_1,\dots,X_n\texttt{)}$, then a predicate of the form $\texttt{:-
dec\_p\_type(p(}\tau_1,\dots,\tau_n\texttt{))}$, where $\tau_1, \dots, \tau_n$
are types, must precede \texttt{p}'s definition. This is interpreted by the
typechecker as $X_i$ is of type $\tau_i$ in \texttt{p}, for all $i \in 1 \dots
n$. See an example in Sect. \ref{machines}.

More details on types in \setlog can be found in the user's manual \cite{Rossi00}.

\subsection{Constraint solving}
As concerns constraint solving, the \setlog solver repeatedly applies specialized
rewriting procedures to its input formula $\Phi$ and returns either $\false$ or
a formula in a simplified form which is guaranteed to be satisfiable with
respect to the intended interpretation. Each rewriting procedure applies a few
non-deterministic rewrite rules which reduce the syntactic complexity of
primitive constraints of one kind. At the core of these procedures is set
unification \cite{Dovier2006}. The execution of the solver is iterated until a
fixpoint is reached, i.e., the formula is irreducible.

The disjunction of formulas returned by the solver represents all the
concrete (or ground) \emph{solutions} of the input formula. Any returned
formula is divided into two parts: the first part is a (possibly empty) list of
equalities of the form $X = t$, where $X$ is a variable occurring in the input
formula and $t$ is a term; and the second part is a (possibly empty) list of
primitive constraints.

\section{\label{usingsetlog}Using \setlog}

In this section we show examples on how \setlog can be used as a programming
language (\ref{programming}) and as an automated theorem prover (\ref{prover}).
The main goal is to provide examples of the
\emph{formula-program duality} enjoined by \setlog code.

\subsection{\label{programming}\setlog as a programming language}

\setlog is primarily a programming language at the intersection of declarative
programming,  set programming \cite{DBLP:books/daglib/0067831} and constraint
programming. Specifically, \setlog is an instance of the general CLP scheme. As
such, \setlog programs are structured as a finite collection of \emph{clauses},
whose bodies can contain both atomic constraints and user-defined predicates.
The following example shows the program side of the formula-program duality of
\setlog code along with the notion of clause.

\begin{example}\label{ex:update}
The following clause corresponds to the Start\_GearExtend event of the Doors
machine of the Event-B model of the LGS. It takes four arguments and the body
is a \setlog formula.
\begin{gather*}
\mathtt{start\_GearExtend}
  (PositionsDG,
   Gear\_ret\_p, Door\_open\_p, Gear\_ret\_p\_) \text{ :-} \\
\quad  Po \In PositionsDG \And{} \\
\quad  \ApplyTo(Gear\_ret\_p,Po,true) \And{} \\
\quad  Door\_open\_p = \Cp(PositionsDG,\{true\}) \And{} \\
\quad  \Foplus(Gear\_ret\_p,Po,false,Gear\_ret\_p\_).
\end{gather*}
%
Note that $Po$ is an existential variable. $\ApplyTo$ and $\Foplus$ are
predicates definable in terms of the atomic constraints provided by $\BR$. For
example, $\ApplyTo$ is as follows:
\begin{gather}
 \ApplyTo(F,X,Y) \text{ :-} \label{eq:applyTo} \\
\quad    F = \{[X,Y]/G\} \And
    [X,Y] \Nin G \And
    \Comp(\{[X,X]\},G,\{\}). \notag
\end{gather}
%

Now we can call $\mathtt{start\_GearExtend}$ by providing inputs and waiting
for outputs:
\[
Pos = \{front,right,left\} \And{} \\
\mathtt{start\_GearExtend}(Pos,
\Cp(Pos,\{true\}),\Cp(Pos,\{true\}),Gear\_ret\_p\_).
\]
returns:
\begin{gather*}
Gear\_ret\_p\_ = \{[front,false]/\Cp(\{right,left\},\{true\})\} \tag*{\qed}
\end{gather*}
\end{example}

As a programming language, \setlog can be used to implement set-based
specifications (e.g., B or Z specifications). As a matter of fact, an
industrial-strength Z specification has been translated into \setlog
\cite{DBLP:journals/jar/CristiaR21b} and students of a course on Z taught both at Rosario
(Argentina) and Parma (Italy) use \setlog as the prototyping language for their
Z specifications \cite{DBLP:journals/corr/abs-2103-14933}. This means that
\setlog can serve as a programming language in which a \emph{prototype} of a
set-based specification can be easily implemented. In a sense, the \setlog
implementation of a set-based specification can be seen as an \emph{executable
specification}.

%

\begin{remark}
A \setlog implementation of a set-based specification is easy to get but
usually it will not meet the typical performance requirements demanded by
users. Hence, we see a \setlog implementation of a set-based  specification
more as a \emph{prototype} than as a final program. On the other hand, given
the similarities between a specification and the corresponding \setlog program,
it is reasonable to think that the prototype is a \emph{correct} implementation
of the specification\footnote{In fact, the translation process can be automated
in many cases.}.

For example, in a set-based specification a table (in a database) with a
primary key is usually modeled as a partial function, $t: X \pfun Y$.
Furthermore, one may specify the update of row $r$ with data $d$ by means of
the \emph{oplus} or \emph{override} ($\oplus$) operator: $t' = t \oplus \{r
\mapsto d\}$. All this can be easily and naturally translated into
\setlog: $t$ is translated as variable $T$ constrained to verify
$\Pfun(T)$, and the update specification is translated as
$\Oplus(T,\{[R,D]\},T\_)$. However, the $\Oplus$ constraint will perform poorly
compared to the \verb+update+ command of SQL, given that $\Oplus$'s
implementation comprises the possibility to operate in a purely logical manner
with it (e.g., it allows to compute $\Oplus(T,\{[a,D]\},\{[a,1],[b,3]\})$ while
\verb+update+ does not). \qed
\end{remark}

Then, we can use these prototypes to make an early validation of the
requirements.
Validating user requirements by means of prototypes entails executing the
prototypes together with the users so they can agree or disagree with the
behavior of the prototypes. This early validation will detect many errors,
ambiguities and incompleteness present in the requirements and possible
misunderstandings or misinterpretations generated by the software engineers.
Without this validation many of these issues would be detected in later stages
of the project thus increasing the project costs.

\subsection{\label{prover}\setlog as an automated theorem prover}

\setlog is also a \emph{satisfiability solver}.  This means that \setlog is a
program that can decide if formulas of some theory are \emph{satisfiable} or
not. In this case the theory is the combination of the decidable (fragments of
the) theories of Fig. \ref{fig:stack}.

Being a satisfiability solver, \setlog can be used as an automated theorem
prover and as a counterexample generator. To prove that formula $\phi$ is a
theorem, \setlog has to be called to prove that $\lnot\phi$ is unsatisfiable.

\begin{example}\label{ex:proofs}
We can prove that set union is commutative by asking \setlog to prove the
following is unsatisfiable:
\begin{gather*}
\Neg(\Cup(A,B,C) \And \Cup(B,A,D) \Implies C = D).
\end{gather*}
\setlog first applies $\Neg$ to the formula, returning:
\begin{gather*}
\Cup(A,B,C) \And \Cup(B,A,D) \And C \Neq D
\end{gather*}
As there are no sets satisfying this formula, \setlog answers $\mathtt{no}$.
Note that the initial formula can also be written as: $\Neg(\Cup(A,B,C)
\Implies \Cup(B,A,C))$. In this case, the result of applying $\Neg$ uses the
$\Ncup$ constraint, $\Cup(A,B,C) \And \Ncup(B,A,C)$. \qed
\end{example}

When \setlog fails to prove that a certain formula is unsatisfiable, it
generates a counterexample. This is useful at early stages of the verification
phase because it helps to find mismatches between the specification and its
properties.

\begin{example}
If the following is run on \setlog:
\begin{gather*}
\Neg(\Cup(A,BBBB,C) \And \Cup(B,A,D) \Implies C = D).
\end{gather*}
the tool will provide the following as the first counterexample:
\begin{gather*}
C = \{\_N3/\_N1\}, \_N3 \Nin D, \Cup(A,\_N2,\_N1), \dots
\end{gather*}
More counterexamples can be obtained interactively.
\qed
\end{example}

\begin{remark}
As we have mentioned in Sect. \ref{formulas}, there are cases where \setlog's
$\Neg$ predicate is not able to correctly compute the negation of its argument.
For instance, if we want to compute the negation of $\ApplyTo$ as defined in
Example \ref{ex:update} with $\Neg$, the result will be wrong---basically due
to the presence of the existentially quantified variable $G$. In that case, the
user has to compute and write down the negation manually.
\end{remark}

Evaluating properties with \setlog helps to run correct simulations by checking
that the starting state is correctly defined. It also helps to \emph{test}
whether or not certain properties are true of the specification or not.
However, by exploiting the ability to use \setlog as a theorem prover, we can
\emph{prove} that these properties are true of the specification. In
particular, \setlog can be used to automatically discharge verification
conditions in the form of state invariants. Precisely, in order to prove that
state transition $T$ (from now on called \emph{operation}) preserves state
invariant $I$ the following proof must be discharged:
\begin{equation}\label{e:inv}
I \land T \implies I'
\end{equation}
where $I'$ corresponds to the invariant in the next state.
If we want to use \setlog to discharge \eqref{e:inv} we have to ask \setlog to
check if the negation of \eqref{e:inv} is \emph{unsatisfiable}. In fact, we
need to execute the following \setlog \emph{query}:
\begin{equation}\label{e:neginv}
\Neg(I \And T \Implies I')
\end{equation}
As we have pointed out in Example \ref{ex:proofs}, \setlog rewrites \eqref{e:neginv} as:
\begin{equation}
I \And T \And \Neg(I')
\end{equation}

\begin{example}\label{ex:invproof}
The following is the \setlog encoding of the state invariant labeled
\texttt{inv2} in machine Doors of the LGS:
\begin{gather*}
\mathtt{doors\_inv2}
  (PositionsDG,
   Gear\_ext\_p, Gear\_ret\_p, Door\_open\_p) \text{ :-} \\
\quad  \Exists(Po \In PositionsDG, \\
\qquad    \ApplyTo(Gear\_ext\_p,Po,false) \And
    \ApplyTo(Gear\_ret\_p,Po,false)
  ) \\
\quad{}    \Implies
      Door\_open\_p = \Cp(PositionsDG,\{true\}).
\end{gather*}
Besides, Rodin generates a proof obligation as \eqref{e:inv} for $\mathtt{start\_GearExtend}$ and $\mathtt{doors\_inv2}$. Then, we can discharge that proof obligation by calling \setlog on its negation:
\[
\Neg(\mathtt{doors\_inv2}
  (PosDG,
   Gear\_ext\_p, Gear\_ret\_p, Door\_open\_p) \And{} \\
\qquad\mathtt{start\_GearExtend}
  (PosDG,
   Gear\_ret\_p, Door\_open\_p, Gear\_ret\_p\_) \\
\qquad\Implies~
\mathtt{doors\_inv2}
  (PosDG,
   Gear\_ext\_p, Gear\_ret\_p\_, Door\_open\_p))
\]
The consequent corresponds to the invariant evaluated in the next state due to
the presence of $Gear\_ret\_p\_$ instead of $Gear\_ret\_p$ in the third
argument---the other arguments are not changed by $\mathtt{start\_GearExtend}$.
\qed
\end{example}

Examples \ref{ex:update} and \ref{ex:invproof} show that \setlog is a
programming and proof platform  exploiting the program-formula duality within
the theories of Fig. \ref{fig:stack}. Indeed, in Example \ref{ex:update}
$\mathtt{start\_GearExtend}$ is treated as a program (because it is executed)
while in Example \ref{ex:invproof} is treated as a formula (because it
participates in a theorem).

\section{\label{encoding}Encoding the Event-B Specification of the LGS in \setlog}

We say `encoding' and not `implementing' the LGS specification because the
resulting \setlog code: \emph{a)} is a  prototype rather than a production
program; and \emph{b)} is a formula as the LGS specification is. In particular,
\setlog provides all the logical, set and relational operators used in the LGS
specification. Furthermore, these operators are not mere imperative
implementations but real mathematical definitions enjoying the formula-program
duality discussed in Sect. \ref{usingsetlog}.

Given that we prove that the \setlog program verifies the
properties proposed by Rodin (see Section \ref{proofs}), we claim the prototype
is a faithful encoding of the Event-B specification.

Due to space considerations we are not going to explain in detail the Event-B
development of Mammar and Laleau. Instead we will provide some examples of how
we have encoded it in \setlog. The interested reader can first take a quick
read to the problem description \cite{DBLP:conf/asm/BoniolW14}, and then
download the Event-B
development\footnote{\url{http://deploy-eprints.ecs.soton.ac.uk/467}} and the
\setlog
program\footnote{\url{http://www.clpset.unipr.it/SETLOG/APPLICATIONS/lgs.zip}}
in order to make a thorough comparison. The Event-B development of the LGS
consists of eleven models organized in a refinement pipeline
\cite{DBLP:journals/sttt/MammarL17}. Each model specifies the behavior and
state invariants of increasingly complex and detailed versions of the LGS. For
example, the first and simplest model specifies the gears of the LGS; the
second one adds the doors that allow the gears to get out of the aircraft; the
third one adds the hydraulic cylinders that either open (extend) or close
(retract) the doors (gears); and so on and so forth. In this development each
Event-B model consists of one Event-B machine.

In the following subsections we show how the main features of the Event-B model
of the LGS are encoded in \setlog.

\subsection{\label{machines}Encoding Event-B machines}

Fig. \ref{fig:b-setlog} depicts at the left the Event-B machine named Gears and
at the right the corresponding \setlog encoding. From here
on, \setlog code is written in \texttt{typewriter} font. We tried to
align as much as possible the \setlog code w.r.t. the corresponding Event-B
code so the reader can compare both descriptions line by line. The \setlog code
corresponding to a given Event-B machine is saved in a file with the name of
the machine (e.g., \verb+gears.pl+).

\newcommand{\q}{\phantom{aa}}
\newcommand{\qq}{\phantom{aaaa}}

\begin{sidewaysfigure}
\begin{multicols}{2}
\textbf{MACHINE} Gears \\
\textbf{SEES} PositionsDoorsGears \\
\textbf{VARIABLES}  gear\_ext\_p \\
\textbf{INVARIANTS} \\
\q  \text{\textsc{inv1}: } $gear\_ext\_p \in PositionsDG \fun BOOL$\\
\textbf{EVENTS} \\
\textbf{Initialisation} \\
\q \textbf{begin} \\
\qq \text{\textsc{act1}: } $gear\_ext\_p := PositionsDG \cross \{TRUE\}$ \\
\q \textbf{end} \\
\textbf{Event} Make\_GearExtended $\defs$ \\
\q  \textbf{any} \\
\qq $po$ \\
\q \textbf{where} \\
\qq \text{\textsc{grd1}: } $po \in PositionsDG \land gear\_ext\_p(po) = FALSE$ \\
\q \textbf{then} \\
\qq \text{\textsc{act1}: } $gear\_ext\_p(po) := TRUE$ \\
\q \textbf{end} \\
\textbf{Event} Start\_GearRetract $\defs$ \\
\q  \textbf{any} \\
\qq $po$ \\
\q \textbf{where} \\
\qq \text{\textsc{grd1}: } $po \in PositionsDG \land gear\_ext\_p(po) = TRUE$ \\
\q \textbf{then} \\
\qq \text{\textsc{act1}: } $gear\_ext\_p(po) := FALSE$ \\
\q \textbf{end} \\
\textbf{END}

\columnbreak

\begin{verbatim}



inv1(PositionsDG,Gear_ext_p) :-
  pfun(Gear_ext_p) & dom(Gear_ext_p,PositionsDG).


init(PositionsDG,Gear_ext_p) :-
  Gear_ext_p = cp(PositionsDG,{true}).

make_GearExtended(PositionsDG,Gear_ext_p,Gear_ext_p_) :-



  Po in PositionsDG &
  applyTo(Gear_ext_p,Po,false) &
  foplus(Gear_ext_p,Po,true,Gear_ext_p_).

start_GearRetract(PositionsDG,Gear_ext_p,Gear_ext_p_) :-



  Po in PositionsDG &
  applyTo(Gear_ext_p,Po,true) &
  foplus(Gear_ext_p,Po,false,Gear_ext_p_).
\end{verbatim}
\end{multicols}
\caption{\label{fig:b-setlog}The Event-B Gears machine at the left and its
\setlog encoding at the right}
\end{sidewaysfigure}

As can be seen in the figure, each invariant, the initialization predicate and
each event are encoded as \setlog clauses. In Event-B the
identifiers for each of these constructs can be any word but in \setlog clause
predicates must begin with a lowercase letter. For instance, event
Make\_GearExtended corresponds to the \setlog clause predicate named
\verb+make_GearExtended+\footnote{Actually, the name of each clause
predicate is prefixed with the name of the machine. Then, is
\texttt{gears\_make\_GearExtended} rather than \texttt{make\_GearExtended}. We
omit the prefix whenever it is clear from context.}.

Each clause predicate receives as many arguments as variables and constants are
used by the corresponding  Event-B construct. For example, \verb+inv1+ waits
for two arguments: \verb+PositionsDG+, corresponding to a set declared in the
Event-B context named PositionsDoorsGears (not shown); and \verb+Gear_ext_p+,
corresponding to the state variable declared in the machine. When an event
changes the state of the machine, the corresponding \setlog clause predicate
contains as many more arguments as variables the event modifies. For example,
Make\_GearExtended modifies variable $gear\_ext\_p$, so
\verb+make_GearExtended+ has \verb+Gear_ext_p_+ as the third argument.
\verb+Gear_ext_p_+ corresponds to the value of \verb+Gear_ext_p+ in the next
state; i.e., the value of $gear\_ext\_p$ after considering the assignment
$gear\_ext\_p(po) := TRUE$. Observe that in \setlog variables must begin with
an uppercase letter and constants with a lowercase letter, while such
restrictions do not apply in Event-B. For example, in the Event-B model we have
the Boolean constant $TRUE$ which in our encoding is written as \verb+true+.
Likewise, in Event-B we have variable $gear\_ext\_p$ which is encoded as
\verb+Gear_ext_p+.

In Fig. \ref{fig:b-setlog} we have omitted type declarations for brevity. We
will include them only when strictly necessary. For instance, the \setlog
clause \verb+inv1+ is actually preceded by its type declaration:
\begin{verbatim}
:- dec_p_type(inv1(stype(positionsdg),stype([positionsdg,bool]))).
\end{verbatim}
where \verb+positionsdg+ is a synonym for \verb+etype([front,right,left])+
and \verb+bool+ is for \verb+etype([true,false])+.


Guards and actions are encoded as \setlog predicates. The identifiers
associated to guards and actions can be provided as comments. \setlog does not
provide language constructs to label formulas. As an alternative, each guard
and action can be encoded as a clause named with the Event-B identifier. These
clauses are then assembled together in clauses encoding events. More on the
encoding of actions in Sect. \ref{actions}.

\subsection{\label{functions}Encoding (partial) functions}
Functions play a central role in Event-B specifications. In Event-B functions
are sets of ordered pairs; i.e., a function is a particular kind of binary
relation. \setlog supports functions as sets of ordered pairs and supports all
the related operators. Here we show how to encode functions and their
operators.

\subsubsection{Function definition.}
In order to encode functions, we use a combination of types and constraints. In
general, we try to encode as much as possible with types, as typechecking
performs better than constraint solving. Hence, to encode $f \in X \pfun Y$, we
declare it as \verb+dec(F,stype([X,Y]))+ and then we assert \verb+pfun(F)+. If
$f$ is a total function, then we conjoin \verb+dom(F,D)+, where \verb+D+ is the
set representation of the domain type.

For instance, predicate \textsc{inv1} in Gears asserts $gear\_ext\_p \in
PositionsDG \fun BOOL$ (Fig. \ref{fig:b-setlog}). In \setlog we declare
\verb+Gear_ext_p+ to be of type \texttt{stype([posi}-\texttt{tionsdg,bool])}
(see the \verb+dec_p_type+ predicate in Sect. \ref{machines}). This type
declaration only ensures $gear\_ext\_p \in PositionsDG \rel BOOL$---i.e., a
binary relation between $PositionsDG$ and $BOOL$. The type assertion is
complemented by a constraint assertion:
 \verb+pfun(Gear_ext_p) & dom(Gear_ext_p,PositionsDG)+, as explained above.


\subsubsection{Function application.}
Function application is encoded by means of the \verb+applyTo+ predicate
defined in \eqref{eq:applyTo}. The encoding is a little bit more general than
Event-B's notion of function application. For instance, in Make\_GearExtended
(Fig. \ref{fig:b-setlog}):
\begin{equation}\label{eq:funcapp}
gear\_ext\_p(po) = FALSE
\end{equation} might be undefined because Event-B's type system
can only ensure $gear\_ext\_p \in PositionsDG \rel BOOL$ and $po \in
PositionsDG$. Actually, the following well-definedness proof obligation is
required by Event-B:
\[
po \in \dom gear\_ext\_p \land gear\_ext\_p \in PositionsDG \pfun BOOL
\]

As said, we encode \eqref{eq:funcapp} as \verb+applyTo(Gear_ext_p,Po,false)+.
\verb+applyTo+ cannot be undefined but it can fail because: \emph{a)} \verb+Po+
does not belong to the domain of \verb+Gear_ext_p+; \emph{b)} \verb+Gear_ext_p+
contains more than one pair whose first component is \verb+Po+; or \emph{c)}
\verb+false+ is not the image of \verb+Po+ in \verb+Gear_ext_p+. Then, by
discharging the proof obligation required by Event-B we are sure that
\verb+applyTo+ will not fail due to \emph{a)} and \emph{b)}. Actually,
$gear\_ext\_p \in PositionsDG \pfun BOOL$ is unnecessarily strong for function
application. As \verb+applyTo+ suggests, $f(x)$ is meaningful when $f$ is
\emph{locally functional} on $x$. For example, $\{x \mapsto 1, y \mapsto 2, y
\mapsto 3\}(x)$ is well-defined in spite that the set is not a function.

\subsubsection{Encoding membership to $\dom$.}
In different parts of the Event-B specification we find predicates such as $ po
\in \dom gear\_ext\_p $. There is a natural way of encoding this in \setlog:
\verb+dom(Gear_ext_p,D) & Po in D+. However, we use an encoding based on the
\verb+ncomp+ constraint because it turns out to be more efficient in
\setlog:
\verb+ncomp({[Po,Po]},Gear_ext_p,{})+. \verb+ncomp+ is the negation of the
\verb+comp+ constraint (composition of binary relations). Then,
\verb+ncomp({[X,X]},F,{})+ states that $\{(X,X)\} \comp F \neq \emptyset$ which
can only hold if there is a pair in $F$ of the form $(X,\_)$. Then, $X$ is in
the domain of $F$.

\subsection{\label{actions}{Encoding action predicates}}
Action predicates describe the state change performed due to an event or the
value of the initial state. Next, we show the encoding of two forms of action
predicates: simple assignment and functional override. In events, the next
state variable is implicitly given by the variable at the left of the action
predicate. In \setlog we must make these variables explicit.

\subsubsection{Encoding simple assignments.}
Consider a simple assignment $x := E$. If this is part of the initialization,
we interpret it as $x = E$; if it is part of an event, we interpret it as $x' =
E$. As we have said in Sect. \ref{machines}, $x'$ is encoded as \verb+X_+.
Hence, simple assignments are basically encoded as equalities to
before-state variables in the case of initialization and to after-state
variables in the case of events.

As an example, part of the initialization of the Gears machine is:
\begin{equation}\label{eq:simpassig}
\text{\textsc{act1}: } gear\_ext\_p := PositionsDG \cross \{TRUE\}
\end{equation}
This is simply encoded in \setlog as follows:
\begin{verbatim}
Gear_ext_p = cp(PositionsDG,{true})
\end{verbatim}

An example of a simple assignment in an event, can be the following (event
ReadDoors, machine Sensors):
\[
\begin{split}
\textsc{act1: }
do&or\_open\_ind := \\
& \{front \mapsto door\_open\_sensor\_valueF, \\
&\ left \mapsto door\_open\_sensor\_valueL, \\
&\ right \mapsto door\_open\_sensor\_valueR\}
\end{split}
\]
Then, we encode it as follows:
\begin{verbatim}
  Door_open_ind_ =
    {[front,Door_open_sensor_valueF],
     [left,Door_open_sensor_valueL],
     [right,Door_open_sensor_valueR]}
\end{verbatim}

\subsubsection{Functional override.}
The functional override $f(x) := E$ is interpreted as $f' = f \oplus \{x
\mapsto E\}$, which in turn is equivalent to $f' = f \setminus \{x \mapsto y |
y \in \ran f\} \cup \{x \mapsto E\}$. This equality is encoded by means of the
\verb+foplus+ constraint:
\begin{verbatim}
foplus(F,X,Y,G) :-
  F = {[X,Z]/H} & [X,Z] nin H & comp({[X,X]},H,{}) & G = {[X,Y]/H}
  or    comp({[X,X]},F,{}) & G = {[X,Y]/F}.
\end{verbatim}
That is, if there is more than one image of \verb+X+ through  \verb+F+,
\verb+foplus+ fails. Then, \verb+foplus(F,X,Y,G)+ is equivalent to:
\[
G = F \setminus \{(X,Z)\} \cup \{(X,Y)\} \text{, for some $Z$}
\]
This is a slight difference w.r.t. the Event-B semantics but it  is correct in
a context where $F$ is intended to be a function (i.e,
$\texttt{pfun(}F\texttt{)}$ is an invariant).

As an example, consider the action part of Make\_GearExtended (Fig. \ref{fig:b-setlog}):
$
gear\_ext\_p(po) := TRUE
$.
As can be seen in the same figure, the encoding is:
\verb+foplus(Gear_ext_p,Po,true,Gear_ext_p_)+.

The following
is a more complex example appearing in event Start\_GearExtend of machine
TimedAspects:
\[
\textsc{act2: }
deadlineGearsRetractingExtending(po) := \\
\qquad\qquad
\{front \mapsto cT+12, left \mapsto cT+16, right \mapsto 16\}(po)
\]
The encoding is the following:
\begin{verbatim}
foplus(DeadlineGearsRetractingExtending,
       Po,M1,DeadlineGearsRetractingExtending_) &
applyTo({[front,M2],[left,M3],[right,16]},Po,M1) &
M2 is CT + 12 & M3 is CT + 16
\end{verbatim}
Given that in \setlog function application is a constraint, we cannot put in
the third argument of \verb+foplus+ an expression encoding $\{front \mapsto
cT+12, \dots\}(po)$. Instead, we have to capture the result of that function
application in a new variable (\verb+M1+) which is used as the third argument.
Along the same lines, it would be wrong to write \verb.CT + 12. in place of
\verb+M2+ in \verb?applyTo({[front,M2],...},Po,M1)? because \setlog (as Prolog)
does not interpret integer expressions unless explicitly
indicated. Precisely, the \verb+is+
constraint forces the evaluation of the integer expression at the right-hand
side.

\subsection{Encoding quantifiers}
The best way of encoding quantifiers in \setlog is by means of the
\verb+foreach+ and \verb+exists+ constraints, which implement RUQ and REQ.
Unrestricted existential quantification is also supported. Then, if an Event-B
universally quantified formula cannot be expressed as a RUQ formula, it cannot
be expressed in \setlog.

\subsubsection{Universal quantifiers.} \label{ssect:quantifiers}
In its simplest form, the RUQ formula $\forall x \in A: \phi$ is written in
\setlog as $\texttt{foreach(} X \texttt{ in } A \texttt{, } \phi \texttt{)}$
\cite[Sect. 5.1]{DBLP:journals/jar/CristiaR21a}\footnote{In turn, the $\Forall$
constraint is defined in \setlog by using RIS and the $\subseteq$ constraint,
by exploiting the equivalence $\forall x \in D: \F(x) \iff D \subseteq \{x \in
D | \F(x)\}$}. However, \verb+foreach+ can also receive four arguments:
$\texttt{foreach(} X \texttt{ in } A \texttt{, [} vars \texttt{], } \phi
\texttt{, } \psi\texttt{)}$, where $\psi$ is a conjunction of \emph{functional
predicates}. Functional predicates play the role of \textsc{let} expressions;
that is, they permit to define a name for an expression \cite[Sect.
6.2]{DBLP:journals/jar/CristiaR21a}. In turn, that name must be one of the
variables listed in $vars$. These variables are implicitly existentially
quantified inside the \verb+foreach+. A typical functional predicate is
\verb+applyTo(F,X,Y)+ because there is only one \verb+Y+ for given \verb+F+ and
\verb+X+. Functional predicates can be part of $\phi$ but in that case its
negation will not always correct.

As an example, consider the second invariant of machine GearsIntermediate:
\[
\begin{split}
\text{\textsc{inv2}: }
\forall po \cdot (& po \in PositionsDG \\ &\implies\lnot(gear\_ext\_p(po)=TRUE \land gear\_ret\_p(po)=TRUE))
\end{split}
\]
This is equivalent to a restricted universal quantified formula:
\begin{equation}\label{eq:ruq}
\begin{split}
\text{\textsc{inv2}: }
\forall po & \in PositionsDG \cdot{} \\
& \lnot(gear\_ext\_p(po)=TRUE \land gear\_ret\_p(po)=TRUE))
\end{split}
\end{equation}
Then, we encode \eqref{eq:ruq} as follows:
\begin{verbatim}
inv2(PositionsDG, Gear_ext_p, Gear_ret_p) :-
  foreach(Po in PositionsDG,[M1,M2],
    neg(M1 = true & M2 = true),
    applyTo(Gear_ext_p,Po,M1) & applyTo(Gear_ret_p,Po,M2)).
\end{verbatim}

In some quantified formulas of Mammar and Laleau's Event-B project
the restricted quantification is not as explicit as in \eqref{eq:ruq}. For
instance, \textsc{inv3} of the Cylinders machine is:
\[
\forall po \cdot door\_cylinder\_locked\_p(po)=TRUE \implies door\_closed\_p(po)=TRUE)
\]
However, given that $po$ must be in the domain of $door\_closed\_p$ then it
must belong to $PositionsDG$.


\subsubsection{Existential quantifiers.}
Like universal quantifiers, also existential quantifiers are encoded by first
rewriting them as REQ, and then by using \setlog's \verb+exists+ constraint.
However, \setlog supports also general (i.e., unrestricted) existential
quantification: in fact, all variables occurring in the body of a clause but
not in its head are implicitly existentially quantified. Thus, the Event-B
construct \textbf{any} is not explicitly encoded. For instance, in the encoding
of event Make\_GearExtended we just state \verb+Po in PositionsDG+ because this
declares \verb+Po+ as an existential variable.

\subsection{Encoding types}
In Event-B type information is sometimes given as membership constraints.
Besides, type information sometimes becomes what can be called \emph{type
invariants}. That is, state invariants that convey what normally is typing
information. As we have said, we try to encode as much as possible with \setlog
types rather than with constraints. Some type invariants can be enforced by the
typechecker. Nevertheless, for the purpose of using this project as a benchmark
for \setlog, we have encoded all type invariants as regular invariants so we
can run all the proof obligations involving them.

\subsubsection{Type guards and type invariants.}
For example, in event
ReadHandleSwitchCircuit (machine Failures) we can find the guard:
\[
Circuit\_pressurized\_sensor\_value \in BOOL
\]
Guards like this are encoded as
actual type declarations and not as constraints:
\begin{verbatim}
dec(Circuit_pressurized_sensor_value,bool)
\end{verbatim}
In this way, the typechecker will reject any attempt to bind this variable to
something different from \verb+true+ and \verb+false+.

As an example of a type invariant we have the following one in machine
HandleSwitchShockAbsorber:
\[
\textsc{inv5: }
Intermediate1 \in BOOL
\]
This is an invariant that can be enforced solely by the typechecker. However,
for the purpose of the empirical comparison, we encoded it as a state
invariant:
\begin{verbatim}
:- dec_p_type(inv5(bool)).
inv5(Intermediate1) :- Intermediate1 in {true,false}.
\end{verbatim}

\subsubsection{Encoding the set of natural numbers ($\nat$).}
The Event-B model includes type invariants involving the set of natural
numbers. This requires to be treated with care because \setlog can only deal
with finite sets so there is no set representing $\nat$. As these invariants
cannot be enforced solely by the typechecker we need to combine types and
constraints. Then, an invariant such as:
\[
\textsc{inv1: } currentTime \in \nat
\]
is encoded by typing the variable as an
integer and then asserting that it is non-negative:
\begin{verbatim}
:- dec_p_type(inv1(int)).
inv1(CurrentTime) :- 0 =< CurrentTime.
\end{verbatim}

There are, however, more complex invariants involving $\nat$ such as:
\[
deadlineUnlockLockDoorsCylinders \in PositionsDG \fun \nat
\]
In this case \setlog's type system can only ensure:
\begin{equation}\label{eq:natfunc}
deadlineUnlockLockDoorsCylinders \in PositionsDG \rel \num
\end{equation}
Then, we need to use constraints to state: \emph{a)}
$deadlineUnlockLockDoorsCylinders$ is a function; \emph{b)} the domain is
$PositionsDG$; and \emph{c)} the range is a subset of $\nat$. The first two
points are explained in Sect. \ref{functions}. The last one is encoded with the
\verb+foreach+ constraint:
\begin{verbatim}
foreach([X,Y] in DeadlineUnlockLockDoorsCylinders, 0 =< Y)
\end{verbatim}
As can be seen, the control term of
the \verb+foreach+ predicate can be an ordered pair \cite[Sect.
6.1]{DBLP:journals/jar/CristiaR21a}. Then, the complete encoding of
\eqref{eq:natfunc} is the following:
\begin{verbatim}
:- dec_p_type(inv5(stype(positionsdg),stype([positionsdg,int))).
inv5(PositionsDG,DeadlineUnlockLockDoorsCylinders) :-
  pfun(DeadlineUnlockLockDoorsCylinders) &
  dom(DeadlineUnlockLockDoorsCylinders,PositionsDG) &
  foreach([X,Y] in DeadlineUnlockLockDoorsCylinders, 0 =< Y).
\end{verbatim}

\section{\label{proofs}Encoding Proof Obligations in \setlog}

Rodin is one of the tools used by Mammar and Leleau in
the project. Rodin automatically generates a set of proof obligations for each
model according to the Event-B verification rules
\cite{DBLP:journals/sttt/AbrialBHHMV10}. Among the many kinds of proof
obligations defined in Event-B, in the case of the LGS project Rodin generates
proof obligations of the following three kinds:
\begin{itemize}
\item Well-definedness (\textsc{wd}).
A \textsc{wd} condition is a predicate describing when an
expression or predicate can be safely evaluated. For instance, the \textsc{wd}
condition for $x \div y$ is $y \neq 0$. Then, if $x \div y$ appears in some
part of the specification, Rodin will generate a proof obligation asking for $y
\neq 0$ to be proved in a certain context.
\item Invariant initialization (\textsc{init}).
Let $I$ be an invariant depending on variable $x$. Let $x := V$ be the
initialization of variable $x$. Then, Rodin generates the following proof
obligation\footnote{This is a simplification of the real situation which,
nonetheless, captures the essence of the problem. All the technical details can
be found in the Event-B literature \cite{DBLP:journals/sttt/AbrialBHHMV10}.}:
\begin{equation}\label{eq:init}
\vdash I[V/x]
\end{equation}
\item Invariant preservation (\textsc{inv}). Let $I$ be as above.
Let $E \defs G \land A$ be an event where $G$ are the preconditions (called
\emph{guards} in Event-B) and $A$ are the postconditions (called \emph{actions}
in Event-B). Say $E$ changes the value of $x$; e.g., there is an assignment
such as $x := V$ in $E$. Then, Rodin generates the following proof obligation:
\begin{equation}\label{eq:inv}
\mathcal{I} \land I \land G \land A \vdash I[V/x]
\end{equation}
where $\mathcal{I} \defs \bigwedge_{j \in J} I_j$ is a conjunction of
invariants in scope other than $I$.
\end{itemize}

Dealing with \textsc{inv} proof obligations deserves some attention. When
confronted with a \emph{mechanized} proof we can think on two general
strategies concerning the hypothesis for that proof:
\begin{itemize}
\item The interactive strategy. When proving a theorem interactively,
the more the hypothesis are available the easiest  the proof. In other words,
users of an interactive theorem prover will be happy to have as many hypothesis
as possible.
\item The automated strategy. When using an automated theorem prover,
some hypothesis can be harmful. Given that automated theorem provers do not
have the intelligence to chose (only) the right hypothesis in each proof step,
they might chose hypothesis that do not lead to the conclusion or that produce
a long proof path. Therefore, a possible working strategy is to run the proof with
just the necessary hypothesis.
\end{itemize}

Our approach is to encode proof obligations by following the `automated
strategy'. For example, in the case of \textsc{inv} proofs, we first try
\eqref{eq:inv} \emph{without} $\mathcal{I}$. If the proof fails, the
counterexample returned by \setlog is analyzed, just the necessary $I_j$ are
added, and the proof is attempted once more. Furthermore, in extreme cases
where a proof is taking too long, parts of $G$ and $A$ are dropped to speed up
the prover. As we further discuss it in Sect. \ref{discussion}, this proof
strategy considerably reduces the need for truly interactive proofs.

The combination between typechecking and constraint solving also helps in
improving proof automation. As we have explained in Sect. \ref{encoding}, the
encoding of \textsc{inv1} in Fig. \ref{fig:b-setlog} does not include
constraints to state that the range of $gear\_ext\_p$ is a subset of $BOOL$.
This is so because this fact is enforced by the typechecker. Hence, the
\setlog encoding of \textsc{inv1} is simpler than \textsc{inv1} itself. As a
consequence, proof obligations involving \textsc{inv1} will be simpler, too.
For example, the \textsc{inv} proof for \textsc{inv1} and event
Make\_GearExtended (of machine Gears) is the following:
\begin{equation}\label{eq:inv1}
\begin{split}
& gear\_ext\_p \in PositionsDG \fun BOOL \land \text{Make\_GearExtended} \\
&\implies gear\_ext\_p' \in PositionsDG \fun BOOL
\end{split}
\end{equation}
However, the \setlog encoding of the negation of \eqref{eq:inv1} is the
following\footnote{Predicate \texttt{inv1} of Fig. \ref{fig:b-setlog} is
expanded to make the point more evident.}:
\begin{verbatim}
neg(pfun(Gear_ext_p) & dom(Gear_ext_p, PositionsDG) &
    make_GearExtended(PositionsDG, Gear_ext_p, Gear_ext_p_)
    implies pfun(Gear_ext_p_) & dom(Gear_ext_p_, PositionsDG)).
\end{verbatim}
That is, \setlog will not have to prove that:
\begin{verbatim}
ran(Gear_ext_p,M) & subset(M,{true,false})
\end{verbatim}
is an invariant because this is ensured by the typechecker.


$PositionsDG$ is a set declared in the context named PositionsDoorsGears where
it is bound to the set $\{front, right, left\}$. $PositionsDG$ is used by the
vast majority of events and thus it participates in the vast majority of proof
obligations. In the \setlog encoding, instead of binding
$PositionsDG$ to that value we leave it free unless it is
strictly necessary for a particular proof.
For example, the proof obligation named ReadDoors/grd/\textsc{wd} (machine
Sensors) cannot be discharged if $left$ is not an element of $PositionsDG$.
Hence, we encoded that proof obligation as follows:
\begin{verbatim}
ReadDoors_grd2 :-
  neg(PositionsDG = {left / M} &
      doors_inv1(PositionsDG, Door_open_p) &
      Door_open_sensor_valueL = true
      implies
        ncomp({[left,left]},Door_open_p,{}) & pfun(Door_open_p)).
\end{verbatim}
Note that: we state the membership of only \verb+left+ to \verb+PositionsDG+;
\textsc{inv1} of machine Doors is needed as an hypothesis; the encoding based
on \verb+ncomp+ is used to state membership to $\dom$; and \verb+pfun+ ensures
that \verb+Door_open_p+ is a function while the type system ensures its domain
and range are correct (not shown).

Furthermore, binding $PositionsDG$ to $\{front, right, left\}$ is crucial in a
handful of proof obligations because otherwise the encoding would fall outside
the decision procedures implemented in \setlog. One
example is passingTime/grd7/\textsc{wd} (machine TimedAspects):
\begin{equation}\label{eq:undec}
\begin{split}
& \ran deadlineOpenCloseDoors \neq \{0\} \\
& \implies \ran(deadlineOpenCloseDoors \nrres \{0\}) \neq \emptyset\\
&\t1 \land (\exists b \in \ran(deadlineOpenCloseDoors \nrres \{0\}) \cdot{} \\
&\t2 (\forall x \ran(deadlineOpenCloseDoors \nrres \{0\}) \cdot b \leq x))
\end{split}
\end{equation}
As can be seen, the quantification domain of both the REQ and RUQ is the same,
the REQ is before the RUQ and the REQ is at the consequent of an
implication. This kind of formulas lies outside the decision procedures
implemented in \setlog unless the quantification domain is a closed set---such
as $\{front, right, left\}$.

\section{\label{discussion}Discussion and Comparison}

The LGS Event-B project of Mammar and Leleau comprises 4.8 KLOC of \LaTeX{}
code which amounts to 213 Kb. The \setlog encoding is 7.8 KLOC long weighting
216 Kb. Although \setlog's LOC are quite more than the encoding of the
specification in \LaTeX{}, we would say both encodings are similar in size---we
tend to use very short lines.  Beyond these numbers, it is worth noting that
several key state variables of the Event-B specification are Boolean functions.
We wonder why the authors used them instead of sets because this choice would
have implied less proof obligations---for instance, many \textsc{wd} proofs
would not be generated simply because function application would be absent.


Table \ref{t:summary} summarizes the verification process carried out with
\setlog\footnote{The verification process was run on a Latitude E7470 with a 4
core Intel(R) Core\texttrademark{} i7-6600U CPU at 2.60GHz with 8 Gb of main
memory, running Linux Ubuntu 18.04.5, SWI-Prolog 7.6.4 and \setlog 4.9.8-10i.}.
Each row shows the proof obligations generated by Rodin and discharged by
\setlog for each refinement level (machine). The meaning of the columns is as
follows: \textsc{po} stands for the total number of proof obligations;
\textsc{init}, \textsc{wd} and \textsc{inv} is the number of each kind of proof
obligations; and \textsc{Time} shows the computing time (in seconds) needed to
discharge those proof obligations (\z means the time is less than one second).
As can be seen, all the 465 proofs are discharged, roughly, in 290 s, meaning
0.6 s in average. Mammar and Leleau \cite[Sect.
10.1]{DBLP:journals/sttt/MammarL17} refer that there are 285 proof obligations,
of which 72\% were automatically discharged. We believe that this number
corresponds to the \textsc{inv} proofs---i.e., those concerning with invariant
preservation. Then, Mammar and Leleau had to work out 80 proof obligations
interactively in spite of Roding using external provers such as Atelier B and
SMT solvers.

\begin{table}
\caption{\label{t:summary}Summary of the verification process}
\centering
\begin{tabularx}{.7\textwidth}{Xrrrrr}
\toprule
\textsc{Machine} & \textsc{po} & \textsc{init} & \textsc{wd} & \textsc{inv} & \textsc{Time} \\\hline
Gears                     &   5 &  1 &  2 &   2 &    \z \\
GearsIntermediate         &  13 &  2 &  5 &   6 &    \z \\
Doors                     &  10 &  2 &  2 &   6 &    \z \\
DoorsIntermediate         &  13 &  2 &  5 &   6 &    \z \\
Cylinders                 &  37 &  5 & 14 &  18 &    \z \\
HandleSwitchShockAbsorber &  29 &  5 &  4 &  20 &    \z \\
ValvesLights              &  12 &  2 &  2 &   8 &    \z \\
Sensors                   &  52 & 12 & 17 &  23 &  55 s \\
TimedAspects              &  98 & 14 & 34 &  50 &   4 s \\
Failures                  &  38 &  9 &  9 &  20 &   3 s \\
PropertyVerification      & 158 & 27 &  1 & 130 & 228 s
  \\\hline
\textsc{Totals}           & 465 & 81 & 95 & 289 & 290 s \\
\bottomrule
\end{tabularx}
\end{table}

The figures obtained with \setlog for the LGS are aligned with previous results
concerning the verification of a \setlog prototype of the Tokeneer ID Station
written from a Z specification \cite{DBLP:journals/jar/CristiaR21b} and the
verification of the Bell-LaPadula security model
\cite{DBLP:journals/jar/CristiaR21}.

There is, however, a proof obligation that, as far as we
understand, cannot be discharged. This proof is
HandleFromIntermediate2ToIntermediate1/\textsc{inv}2 of the TimedAspects
machine which would prove that event HandleFromIntermediate2ToIntermediate1
preserves $0 \leq deadlineSwitch$. However, the assignment $deadlineSwitch :=
currentTime+(8-(2/3)*(deadlineSwitch-currentTime))$, present in that event,
implies that the invariant is preserved iff $deadlineSwitch \leq
(5/2)*currentTime + 12$. But there is no invariant implying that inequality.
Both $deadlineSwitch$ and $currentTime$ are first declared in the TimedAspects
machine which states only two invariants concerning these variables: both of
them must be non-negative integers.

Mammar and Leleau use ProB and AnimB besides Rodin during the verification
process. They use these other tools in the first refinement steps to try to
find obvious errors before attempting any serious proofs. For instance, they
use ProB to check that invariants are not trivially falsified. In this sense,
the tool is used to find counterexamples for invalid invariants. In general,
ProB cannot prove that an invariant holds, it can only prove that it does not
hold---as far as we know ProB does not implement a decision procedure for a
significant fragment of the set theory underlying Event-B. \setlog could
potentially be used as a back-end system to run the checks carried out by
Mammar and Leleau. It should produce more accurate and reliable results as it
implements several decision procedures as stated in Sect. \ref{setlog}.

Nevertheless, the main point we would like to discuss is our approach to
automated proof. That is, the automated strategy mentioned in Sect.
\ref{proofs} plus some details of the \setlog encoding of the LGS. Our approach
is based on the idea expressed as \emph{specify for automated proof}. In other
words, when confronted with the choice between two or more ways of specifying a
given requirement, we try to chose the one that improves the chances for
automated proof. Sooner or later, automated proof hits a computational
complexity wall that makes progress extremely difficult. However, there are
language constructs that move that wall further away than others.

For example, the encoding of function application by means of \verb+applyTo+,
the encoding of functional override by means of \verb+foplus+, and the encoding
of membership to $\dom$ by means of \verb+ncomp+, considerably simplify
automated proofs because these are significantly simpler than encodings based
on other constraints.

However, in our experience, the so-called automated strategy provides the
greatest gains regarding automated proof. As we have said, we first try to
discharge a proof such as \eqref{eq:inv} without $\mathcal{I}$. If the proof
fails we analyze the counterexample returned by \setlog and add a suitable
hypothesis---i.e., we pick the right $I_j \in \mathcal{I}$. This process is
iterated until the proof succeeds. Clearly, this proof strategy requires some
degree of interactivity during the verification process. The question is, then,
whether or not this approach is better than attempting to prove \eqref{eq:inv}
as it is and if it does not succeed, an interactive proof assistant is called
in. Is it simpler and faster our strategy than a truly interactive proof?

We still do not have strong evidence to give a conclusive answer, although we
can provide some data. We have developed a prototype of an interactive proof
tool based on the automated strategy \cite{10.1093/comjnl/bxab030}. Those
results provide evidence for a positive answer to that question. Now, the data
on the LGS project further contributes in the same direction. In this project,
\setlog discharges roughly 60\% of the proofs of Table \ref{t:summary} in the
first attempt. In the vast majority of the remaining proofs only \emph{one}
evident $I_j \in \mathcal{I}$ is needed. For example, in many proofs we had to
add the invariant $0 \leq currentTime$ as an hypothesis; and in many others an
invariant stating that some variable is a function.



In part, our approach is feasible because it is clear what \setlog can prove
and what it cannot, because it implements decision procedures. As
a matter of fact, the proof of \eqref{eq:undec} is a good example about the
value of working with decision procedures: users can foresee the behavior of
the tool and, if possible, they can take steps to avoid undesired behaviors.

\section{\label{concl}Final Remarks}

This paper provides evidence that many B specifications can be easily
translated into \setlog. This means that \setlog can serve as a
\emph{programming language} in which \emph{prototypes} of those specifications
can be immediately implemented. Then, \setlog itself can be used to
automatically prove or disprove that the specifications verify the proof
obligations generated by tools such as Rodin. In the case study presented in
this paper, \setlog was able to discharge all such proof obligations.

In turn, this provides evidence that CLP and set theory are valuable tools
concerning formal specification, formal verification and prototyping. Indeed
\setlog, as a CLP instance, enjoys properties that are hard to find elsewhere.
In particular, \setlog code can be seen as a program but also as a set formula.
This duality allows to use \setlog as both, a programming language and an
automated verifier for its own programs. In \setlog, users do not need to
switch back and forth between programs and specifications: programs \emph{are}
specifications and specifications \emph{are} programs.

\bibliographystyle{splncs}
\bibliography{/home/mcristia/escritos/biblio.bib}

\end{document}